\documentclass[sigconf]{acmart}

\usepackage{blindtext}
\usepackage{xcolor}
\usepackage{graphicx}
\usepackage{subcaption}
\usepackage{booktabs}
\usepackage{tcolorbox}
\usepackage{blindtext}
\usepackage{amsmath}
\usepackage{tcolorbox}
\usepackage{listings}
\usepackage{soul}
\usepackage{xcolor}

\usepackage{tikz}
\usetikzlibrary{positioning, arrows.meta, shapes.geometric, fit, backgrounds, calc}

\definecolor{viridis1}{RGB}{68,1,84}
\definecolor{viridis2}{RGB}{59,82,139}
\definecolor{viridis3}{RGB}{33,145,140}
\definecolor{viridis4}{RGB}{94,201,98}
\definecolor{viridis5}{RGB}{204,180,0} 
\definecolor{viridis6}{RGB}{0,0,0} 

\newcommand{\highlightOne}[1]{\sethlcolor{viridis1!50}\hl{#1}}
\newcommand{\highlightTwo}[1]{\sethlcolor{viridis2!50}\hl{#1}}
\newcommand{\highlightThree}[1]{\sethlcolor{viridis3!50}\hl{#1}}
\newcommand{\highlightFour}[1]{\sethlcolor{viridis4!50}\hl{#1}}
\newcommand{\highlightFive}[1]{\sethlcolor{viridis5!50}\hl{#1}}
\newcommand{\highlightSix}[1]{\sethlcolor{viridis6!20}\hl{#1}}

\lstdefinestyle{highlightedJSON}{
    basicstyle=\ttfamily\tiny,
    breaklines=true,
    escapeinside={(*@}{@*)}, 
    literate={\{}{{\{}}1
             {\}}{{\}}}1
             {,}{{,}}1
             {:}{{:}}1
             {[}{{[}}1
             {]}{{]}}1,
}

\newcommand{\code}[1]{%
  {\setlength{\fboxsep}{1.5pt}
  \setlength{\fboxrule}{0pt}
  \raisebox{0pt}[.5\height][0.1\depth]{
    \fbox{\colorbox{gray!20}{\texttt{#1}}}}%
  }%
}

\AtBeginDocument{%
  }

\setcopyright{none}
\copyrightyear{2025}
\acmYear{2025}
\acmDOI{XXXXXXX.XXXXXXX}

\setcopyright{rightsretained}
\acmConference[ITiCSE 2025 (to appear in)]{Proceedings of the 2024 Innovation and Technology in Computer Science Education}{June 30--July 2, 2025}{Nijmegen, The Netherlands}
\acmBooktitle{Proceedings of the 2025 Innovation and Technology in Computer Science Education (to appear in ITiCSE 2025), June 30--July2, 2025, Nijmegen, The Netherlands}
\acmISBN{978-1-4503-XXXX-X/18/06}

\begin{document}

\title[Counting the Trees in the Forest]{Counting the Trees in the Forest: Evaluating Prompt Segmentation for Classifying Code Comprehension Level}

\author{David H. Smith IV}
\email{dhsmith2@illinois.edu}
\orcid{0000-0002-6572-4347}
\affiliation{%
  \institution{University of Illinois}
  \city{Urbana}
  \country{USA}
}

\author{Max Fowler}
\email{mfowler5@illinois.edu}
\orcid{0000-0002-4730-447X}
\affiliation{%
  \institution{University of Illinois}
  \city{Urbana}
  \country{USA}
}

\author{Paul Denny}
\email{p.denny@auckland.ac.nz}
\orcid{0000-0002-5150-9806}
\affiliation{%
  \institution{University of Auckland}
  \country{New Zealand}
}

\author{Craig Zilles}
\email{zilles@illinois.edu}
\orcid{0000-0003-4601-4398}
\affiliation{%
  \institution{University of Illinois}
  \city{Urbana}
  \country{USA}
}

\renewcommand{\shortauthors}{D.H. Smith IV, M. Fowler, P. Denny, C. Zilles}

\begin{abstract}
Reading and understanding code are fundamental skills for novice programmers, and especially important with the growing prevalence of AI-generated code and the need to evaluate its accuracy and reliability.  ``Explain in Plain English'' questions are a widely used approach for assessing code comprehension, but providing automated feedback, particularly on comprehension levels, is a challenging task.  This paper introduces a novel method for automatically assessing the comprehension level of responses to ``Explain in Plain English'' questions. Central to this is the ability to distinguish between two response types: multi-structural, where students describe the code line-by-line, and relational, where they explain the code's overall purpose. Using a Large Language Model (LLM) to segment both the student's description and the code, we aim to determine whether the student describes each line individually (many segments) or the code as a whole (fewer segments). We evaluate this approach's effectiveness by comparing segmentation results with human classifications, achieving substantial agreement.  We conclude with how this approach, which we release as an open source Python package, could be used as a formative feedback mechanism.   
\end{abstract}


\maketitle

\section{Introduction}

The ability to comprehend and articulate the purpose of code---though always
considered a vital skill for novice programmers~\cite{lopez2008relationships,
xie2019theory}---has gained new significance in the context of Human-GenAI
collaborative coding. A classic activity for assessing code comprehension is
the Explain-in-Plain-English (EiPE) question, where students articulate the
purpose of a given code snippet—typically a single function—in natural
language. While these questions are effective in assessing
comprehension~\cite{lopez2008relationships, venables2009closer}, they are
time-intensive to grade manually~\cite{fowler2021should} and require a high
degree of authoring overhead to label data for training auto-grading
models~\cite{fowler2021autograding}.

To reduce the authoring overhead of these autograders, \citet{smith2024code}
proposed a code generation-based grading approach. In this method, a student's
response is used to generate code via an LLM, and a suite of unit tests is applied to
verify whether the code generated from the response is functionally equivalent
to the original code that the student was asked to explain. While this approach
effectively distinguishes functionally correct responses from incorrect ones,
it is limited in that it does not differentiate between \textit{high-level}
responses, which describe the code's purpose, and \textit{low-level} responses,
which focus on its implementation details~\cite{denny2024explaining,
smith2024prompting}. This distinction is critical for assessing code
comprehension.

In the spirit of the code generation based grading approach---namely through
enabling auto-grading approaches that require minimal authoring burden---we
propose a novel approach to automatically assess \textit{comprehension level}
for responses to EiPE questions. Here, we leverage an LLM to take a student's response to an EiPE
question, segment the response, and map the segments of that response to the lines of code that they
describe. The intuition behind this approach being that if a student provides a
high-level description it should map to the code as a whole. However, if they
provide a low-level description, this should result in multiple segments where
each maps to a different line or group of lines in the code. To evaluate this
approach we address the following research questions:
\vspace{-1em}
\begin{enumerate}
  \item[\hspace{2em}\textbf{RQ1:}] How does classification as \textit{high} and
    \textit{low} level using the proposed approach compare to human labeling?
  \item[\hspace{2em}\textbf{RQ2:}] What post-processing steps can be applied to
    improve the performance of the segmentation approach?
\end{enumerate}
We release this approach as a feature in 
\texttt{eiplgrader}\footnote{https://github.com/CoffeePoweredComputers/eiplgrader} package, an open-source
Python package for autograding EiPE questions.

\section{Background}

\begin{figure*}
\centering
\begin{subfigure}{\textwidth}
\centering
\begin{tikzpicture}[node distance=5cm, auto]
\tikzstyle{promptBox} = [rectangle, draw, align=left, rounded corners, inner sep=5pt]
\pgfdeclarelayer{foreground}
\pgfdeclarelayer{background}
\pgfsetlayers{background,main,foreground}
\begin{pgfonlayer}{background}
\node (system) [promptBox, font=\footnotesize, text width=0.95\textwidth, inner sep=10pt] {
    \textbf{Task:} Create a one-to-one mapping between each segment of a given
    explanation and the group of lines in the given code which that phrase is
    associated with. Not all of the description needs to be used. Not all of the
    code needs to be used. It is very important to only use the words in the user's
    provided explanation. One segment can map to multiple lines.
    Here is the code:
    \begin{lstlisting}[
        language=C, 
        basicstyle=\footnotesize\ttfamily, 
        keywordstyle=\bfseries\color{purple}, 
        commentstyle=\itshape\color{green}
    ]
int sumOfPositives(int arr[], int size) {
    int x = 0;
    for (int i = 0; i < size; i++) {
        if (arr[i] > 0) {
            x += arr[i];
        }
    }
    return x;
}\end{lstlisting}
};
\end{pgfonlayer}

\begin{pgfonlayer}{foreground}
\node[rectangle, fill=black, text=white, font=\footnotesize, anchor=north west, rounded corners] at (-8.6, 2.5) {\textbf{System}};
\end{pgfonlayer}
\end{tikzpicture}
\caption{System prompt providing the instructions for the segmentation task and the code to be used.}
\label{fig:system-prompt}
\end{subfigure}

\vspace{0.5em}

\begin{subfigure}{0.48\textwidth}
\centering
\begin{tikzpicture}[node distance=5cm, auto]
\pgfdeclarelayer{foreground}
\pgfdeclarelayer{background}
\pgfsetlayers{background,main,foreground}
\begin{pgfonlayer}{background}
\tikzstyle{promptBox} = [rectangle, draw, align=left, rounded corners]

\node (user) [promptBox, font=\footnotesize, text width=0.9\columnwidth,  align=left, inner sep=10pt] {
\textbf{Explanation:} \highlightSix{input is values with array and length}. \highlightOne{initially set x to zero}, and \highlightTwo{use for loop to set start i from zero and smaller than length, increasing by 1 for i each run}. If \highlightThree{values are bigger than zero}, then \highlightFour{x plus equal values}. \highlightFive{it will return to x at end}.
};

\node (assistant) [promptBox, text width=0.9\columnwidth, below of=user, inner xsep=10pt] {
\begin{lstlisting}[style=highlightedJSON]
{
  "groups": [
    {
      "code": "int sumOfPositives(int arr[], int size) {",
      "explanation_portion": "(*@\highlightSix{input is values with array and length}@*)"
    },
    {
      "code": "int x = 0;",
      "explanation_portion": "(*@\highlightOne{initially set x to zero}@*)"
    },
    {
      "code": "for (int i = 0; i < size; i++) {",
      "explanation_portion": "(*@\highlightTwo{use for loop to set start i from zero and smaller than length, increasing by 1 for i each run}@*)"
    },
    {
      "code": "if (arr[i] > 0) {",
      "explanation_portion": "(*@\highlightThree{values are bigger than zero}@*)"
    },
    {
      "code": "x += arr[i];",
      "explanation_portion": "(*@\highlightFour{x plus equal values}@*)"
    },
    {
      "code": "return x;",
      "explanation_portion": "(*@\highlightFive{it will return to x at end}@*)"
    }
  ]
}\end{lstlisting}
};

\draw[->] (user.south) -- (assistant.north);
\end{pgfonlayer}

\begin{pgfonlayer}{foreground}
\node[rectangle, fill=black, text=white, font=\footnotesize, anchor=north west, rounded corners] at (-4, 1.1) {\textbf{User}};
\node[rectangle, fill=black, text=white, font=\footnotesize, anchor=north west, rounded corners] at (-4, -1) {\textbf{Assistant}};
\end{pgfonlayer}

\end{tikzpicture}
\caption{Explanation and Response Pair 1 -- Multi-structural}
\label{fig:multistructural}
\end{subfigure}
\hfill
\begin{subfigure}{0.48\textwidth}
\centering
\begin{tikzpicture}[node distance=7.05cm, auto]

\tikzstyle{promptBox} = [rectangle, draw, align=center, rounded corners]

\pgfdeclarelayer{foreground}
\pgfdeclarelayer{background}
\pgfsetlayers{background,main,foreground}
\begin{pgfonlayer}{background}

\node (user2) [promptBox, font=\footnotesize, text width=0.9\columnwidth, align=left, inner sep=10pt]{
\textbf{Explanation:} \highlightOne{sums all positive numbers in the array}.
};

\node (assistant2) [promptBox, text width=0.9\columnwidth, below of=user, inner xsep=10pt] {
\begin{lstlisting}[style=highlightedJSON]
{
  "groups": [
    {
      "code": "int sumOfPositives(int arr[], int size) {"
             "    int x = 0;"
             "    for (int i = 0; i < size; i++) {"
             "        if (arr[i] > 0) {"
             "            x += arr[i];"
             "        }"
             "    }"
             "    return x;"
             "}",
      "explanation_portion": "(*@\highlightOne{sums all positive numbers in the array}@*)"
    }
  ]
}\end{lstlisting}
};

\draw[->] (user2.south) -- (assistant2.north);
\end{pgfonlayer}

\begin{pgfonlayer}{foreground}
\node[rectangle, fill=black, text=white, font=\footnotesize, anchor=north west, rounded corners] at (-4, 0.65) {\textbf{User}};
\node[rectangle, fill=black, text=white, font=\footnotesize, anchor=north west, rounded corners] at (-4, -4.65) {\textbf{Assistant}};
\end{pgfonlayer}

\end{tikzpicture}
\caption{Explanation and Response Pair 2 -- Relational}
\label{fig:relational}
\end{subfigure}

\vspace{-0.5em}
\caption{The Components of the Prompt Used for the Segmentation Approach Used in this Study.}
\label{fig:comparison}
\vspace{-1em}
\end{figure*}

The related work for this study is organized into three areas.
Section~\ref{sec:code-comprehension} covers foundational programming skills like
tracing, explaining, and writing code, focusing on Explain in Plain English
(EiPE) questions as tools for assessing code comprehension.
Section~\ref{sec:autograding} examines the challenges of grading EiPE responses
and summarizes autograding methods, noting their strengths and limitations.
Section~\ref{sec:nlp} provides an overview of natural language programming and
prompting, highlighting efforts to integrate prompt engineering into computing
education and the gaps our approach addresses.

\subsection{Comprehension as a Foundational Skill}\label{sec:code-comprehension}

Learning to program requires the development of a range of inter-related skills,
including tracing, explaining, and writing code~\cite{lister2009further,
lopez2008relationships}. Tracing provides an important foundation, enabling
students to understand program flow by simulating execution step-by-step, and
explaining builds on this by requiring students to abstract and articulate the
purpose of code, moving from a line-by-line analysis to a broader conceptual
understanding. Research indicates that tracing and explaining code often precede
and support the development of code-writing abilities, highlighting their
importance in the learning process~\cite{venables2009closer,
fowler2022reevaluating}.

One popular approach for assessing code reading and understanding is the
``Explain in Plain English'' (EiPE) question type.  First described by
\citet{whalley2006australasian}, EiPE questions were used to assess the
comprehension skills of novice programmers, using the SOLO taxonomy as a
framework for categorizing and analyzing the natural language responses.  Unlike
tracing, which focuses on predicting the behavior of individual lines, EiPE
tasks require students to articulate the purpose of code segments in natural
language, emphasizing abstract reasoning ~\cite{murphy2012explain}. 

Despite their pedagogical value, grading EiPE responses has traditionally been
time-intensive and subjective. In fact, a survey study by
\citet{fowler2021should} found that consistent and scalable scoring is one of
the primary barriers to the adoption of EiPE questions in programming education,
with many educators citing challenges related to the variability in student
responses and the effort required to ensure grading fairness.  This has led to
interest in more scalable approaches for assessing and providing feedback on
EiPE tasks~\cite{fowler2021autograding, chen2022peer}.

\subsection{Grading Explain-in-Plain-English}\label{sec:autograding}

Although the SOLO taxonomy is often used for classifying answers to EiPE questions, it is often difficult to apply given the need to assess not just abstraction, but also correctness and ambiguity~\cite{chen2020validated}. Achieving consistent and scalable scoring is a well-documented challenge given the subjectivity of grading natural language responses~\cite{fowler2021should}. To address this, several automated grading approaches have been developed. One notable solution involves a logistic classifier trained on human-labeled data to evaluate EiPE responses~\cite{fowler2021autograding}. While this method achieves grading accuracy comparable to that of human teaching assistants, it requires a substantial upfront investment in labeled datasets for each question.
Additionally, given the dicotomous nature of the grading mechanism, it is unable to provide actionable feedback to students on where there response had shortcomings.

More recently, Smith and Zilles proposed Code Generation Based Grading (CGBG) which leverages large language models (LLMs) to generate code from a student's natural language response \cite{smith2024code}. The generated code is then validated using a suite of instructor-defined unit tests to ensure it is functionally equivalent to the code being explained by the student.  This approach removes the need for extensive pre-labeled datasets and provides novel kinds of feedback, including displaying the generated code and test results to students~\cite{smith2024prompting, kerslake2024integrating}. Additionally, recent evaluations have highlighted the potential for using CGBG with in multiple natural languages, thus providing support for non-native English speaking student cohorts \cite{smith2024explain}. This has lead the authors to suggest that, by using this grading approach, the question type is transformed from an ``Explain in Plain English'' question to an ``Explain in Plain Language'' question.

However, while CGBG provides objective feedback by executing the generated code, it does not differentiate between surface-level (multi-structural) and deeper (relational) responses~\cite{smith2024code, denny2024explaining}. For example, a multi-structural response might describe the behavior of individual parts of the code without identifying their collective purpose, yet would be graded by CGBG as equally correct compared to a relational response that does describe the abstract goal. This distinction is important for identifying students' comprehension levels and guiding them towards higher-order reasoning.  The inability of current autograding systems to assess these cognitive distinctions limits their effectiveness \cite{denny2024explaining}.

\vspace{-0.1cm}
\subsection{Natural Language Programming}\label{sec:nlp}

Writing natural language explanations of code, as a way to demonstrate code comprehension, is closely related to the idea of natural language programming.  Natural language programming involves specifying solutions to computational tasks using natural language---a concept that has been controversial.  Dijkstra famously criticized the idea, arguing that the ambiguity of natural language makes it poorly-suited for precise computational tasks~\cite{dijkstra2005foolishness}. Despite these concerns, recent advancements in large language models (LLMs) have reignited interest in this area, though the translation of abstract ideas into concrete code remains challenging~\cite{xu2022inide}.

One critical skill for effective use of LLMs to generate code is the ability to craft clear and precise prompts~\cite{denny2023conversing}.  Many students find this process difficult, and yet it is emerging as an important new skill \cite{denny2024cacm}. Recent work has explored integrating prompting activities into introductory programming courses, to provide opportunities for students to practice constructing prompts and evaluating the resulting code~\cite{kerslake2024integrating}.  One particular approach, ``Prompt Problems'', focuses on helping students write prompts that generate functionally correct code \cite{denny2024prompt}. Similar to CGBG, Prompt Problems rely on LLMs to generate code from student prompts, using test cases to evaluate correctness.

Prior work on teaching prompting has primarily focused on functional correctness, with little consideration for aspects of code quality such as style, readability, or efficiency.  The work we describe here builds on these efforts, aiming to extend the scope of evaluation beyond correctness to include feedback on the level of comprehension. 

\section{Segmentation Approach}

Our segmentation approach---which is done on a per-question basis---is designed
to analyze a student's response and map it to the specific line(s) of code it
describes. This process begins by providing the large language model (LLM) with
a \textit{system prompt} (Figure~\ref{fig:system-prompt}). The system prompt
contains three key components: 1) a description of the task it is performing, 2)
some guidelines on how it is to perform the task, and 3) the code snippet that
the student is to describe.

To further tune the model's performance, we provide the model with two,
few-shot examples: 1) an example of a mapping for a multi-structural response
(Figure~\ref{fig:multistructural}) and 2) an example of a mapping for a
relational response (Figure~\ref{fig:relational}). These examples are used to
help the model understand the difference between the two types of responses and
how they should be segmented and mapped to the code. To guarantee adherence to
the desired segmentation format in JSON, we leverage OpenAI's GPT-4o with
structured output capabilities. This feature allows the model to produce
well-formed JSON output that meets the specified
requirements\footnote{\url{https://openai.com/index/introducing-structured-outputs-in-the-api/}}.

With the system prompt and several few-shot examples in place, the model is then
provided with a student's response to the given question which it segments and
maps to the code snippet. The model outputs the resulting JSON which is then
used for the classification of the response as either multi-structural or
relational.

\section{Methods}

The data for this study was sourced from a large introductory programming
course at a major research university in Australasia. This course, mandatory
for all students in the Faculty of Engineering, is taken during their first
year of study and covers standard CS1 topics. At the time of data collection,
889 students were enrolled, with 841 participating. Access to this data was
approved by the university's Human Participants Ethics Committee under protocol
number UAHPEC25279.

\subsection{Exercises}

\begin{figure}
  \centering
  \includegraphics[width=0.95\columnwidth]{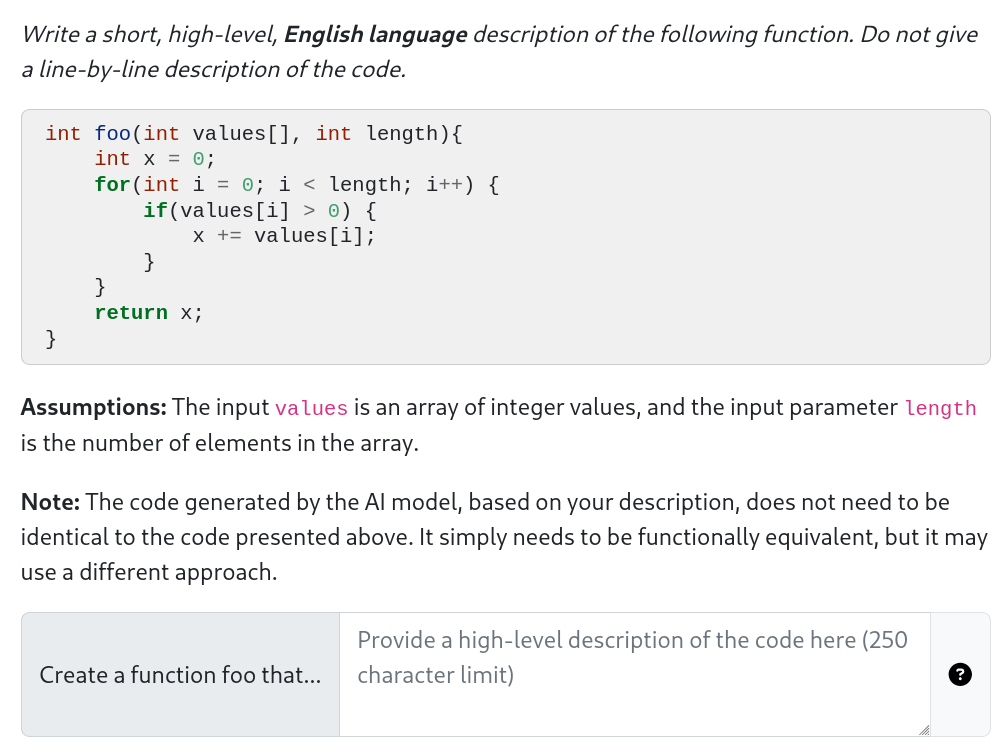}
  \caption{Question interface for A-Q3: Index of last zero.}
  \label{fig:pl-indexlastzero}
\end{figure}

\begin{table}[htbp]
    \caption{Questions for Lab A (Loops, Arrays and Functions) and Lab B (Strings, Text Processing and 2D arrays)}\label{tab:lab_questions}
    \centering
    \begin{tabular}{ccc}
      \hline
      \textbf{Lab} & \textbf{Question} & \textbf{Description} \\ \hline
      A & Q1 & Sum between a and b inclusive \\ 
      A & Q2 & Count even numbers in array \\ 
      A & Q3 & Index of last zero (see Figure \ref{fig:pl-indexlastzero}) \\ 
      A & Q4 & Sum positive values \\ \hline
      B & Q1 & Reverse a string \\ 
      B & Q2 & Calculate sum of row in 2D array \\ 
      B & Q3 & Is a vowel contained in a string? \\ 
      B & Q4 & Does a string contain a substring? \\ \hline
    \end{tabular}
\end{table}

Students completed a total of 8 exercises across two labs. Each lab contained
four exercises. The exercises were designed to cover a range of topics and
difficulty levels. The exercises for Lab A (Loops, Arrays and Functions) and
Lab B (Strings, Text Processing and 2D arrays) are shown in
Table~\ref{tab:lab_questions}. To administer these questions we used the
PrairieLearn online-assessment platform~\cite{west2015prairielearn}. 
These activities were given in a formative environment where students were given the
opportunity to submit multiple times with a maximum of 20 attempts for each
question. The questions were autograded using the approach introduced by
\citet{smith2024code}. Students were given feedback on their responses after
each submission in the form of the code generated from their response and the
results of test cases run on that code.

\subsection{Human Labeling}

To establish a data set for evaluating the segmentation-based classification of
responses, two researchers---using the modified Structure of Observed Learning
Outcome (SOLO) taxonomy~\cite{biggs2014evaluating} introduced by
\citet{clear2008reliably}---independently classified a random subset of 200
responses for each question. Beginning with questions A-Q1 and B-Q4, the
researchers classified the responses as either \textit{multi-structural
correct}, \textit{relational correct}, or \textit{incorrect}. The responses to
both questions were categorized and achieved a Cohen's
$\kappa$~\cite{cohen1960coefficient} of 0.80 and 0.83, respectively, indicating
a substantial agreement~\cite{landis1977measurement}. Following reconciliation,
the remaining responses were classified individually with each of the
researchers classifying half of the responses for each of the remaining
questions.

\section{Results}

\begin{figure}
  \includegraphics[width=0.95\columnwidth]{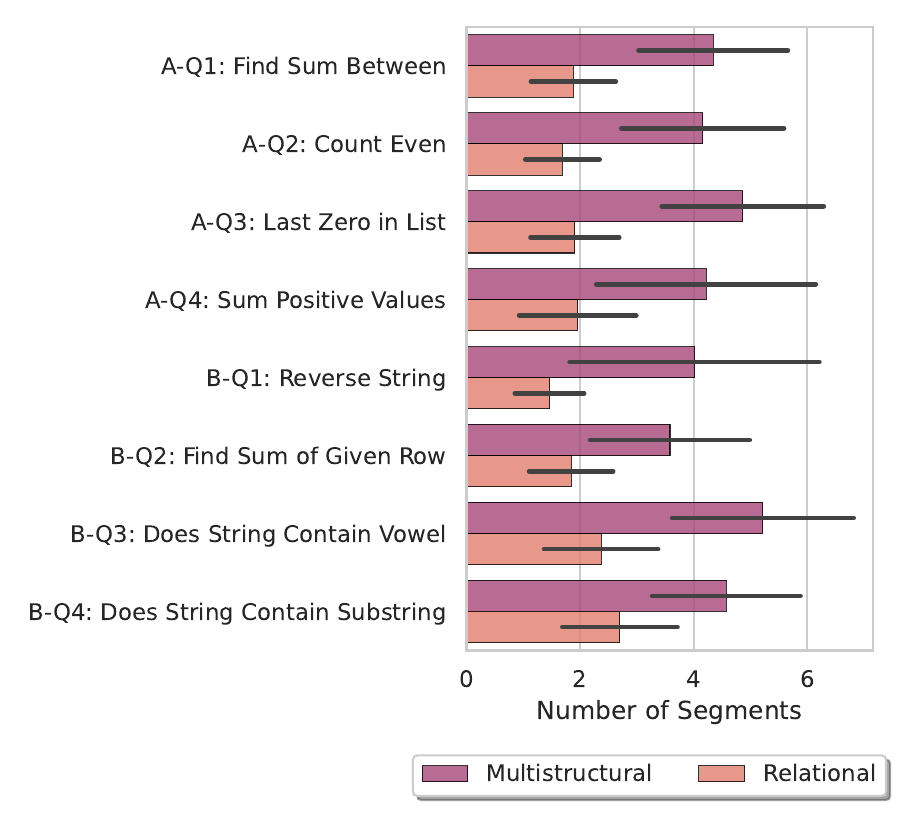}
  \caption{The mean and stdev in the number of segments generated
  for multistructural and relational responses.}
  \label{fig:segmented}
\end{figure}


\begin{figure*}
  \centering
  \begin{subfigure}{0.45\textwidth}
    \includegraphics[width=\textwidth]{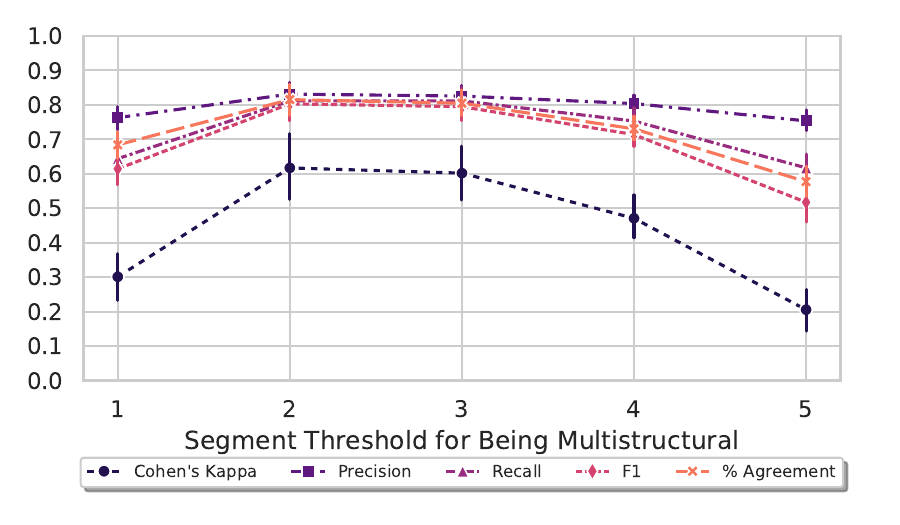}
    \caption{Performance of the segmentation approach for classifying student responses.}
    \label{fig:segmentation-performance}
  \end{subfigure}
  \hfill
  \begin{subfigure}{0.45\textwidth}
    \includegraphics[width=\textwidth]{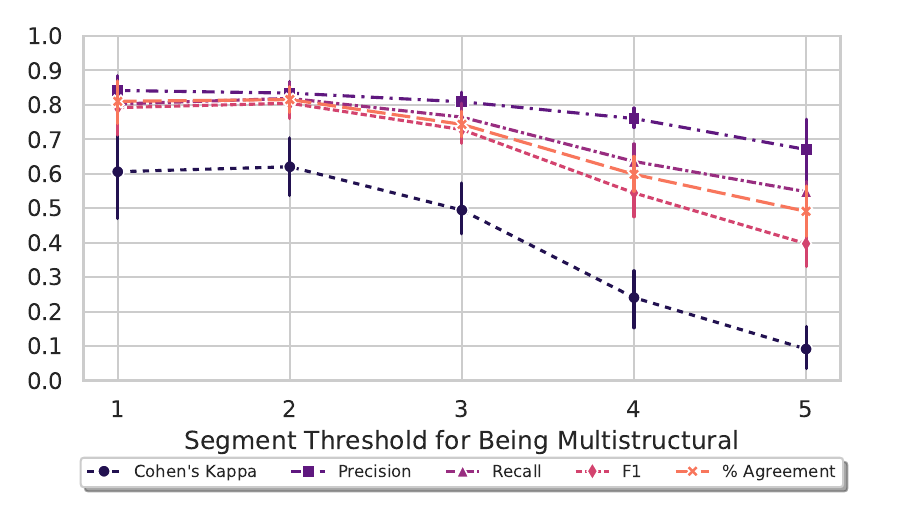}
    \caption{Performance of the segmentation after function definition removal post-processing.}
    \label{fig:segmentation-performance-no-definitions}
  \end{subfigure}
  \vspace{-0.25cm}
  \caption{Performance of the segmentation approach for classifying student responses. Responses containing a number of segments above each of given thresholds is classified as multistructural and  those at or below are classified as  relational.} \label{fig:segmentation-performance-comparison}
  \vspace{-0.2cm}
\end{figure*}


To evaluate differences in segmentation outcomes, we compared the number of
segments generated for responses labeled as \textit{multi-structural} versus
\textit{relational}. Figure~\ref{fig:segmented} shows that
\textit{multi-structural} responses consistently had more segments, aligning
with our expectations and the model's fine-tuning examples. However, it is worth
noting that the standard deviations for the segmentation counts is reasonably
high, meaning there is a significant overlap between the distributions which
places an upper limit on the degree to which said distributions can be
separated. However, the overall results suggest that the segmentation process
captures some of the distinctions between \textit{relational} and
\textit{multi-structural responses}.

\subsection{Performance of Segmentation Classification}

Using the resulting segmentation groups, we investigated how varying
segmentation thresholds affected the performance of the segmentation-based
classification approach. The thresholds ranged from 1 to 4, where each threshold
represents the number of segments above which a response is classified as
multi-structural. To evaluate the approach, we calculated Cohen's $\kappa$, F1
score, precision, recall, and percent agreement for each threshold.

\paragraph{\textbf{Initial Segmentation}}

Figure~\ref{fig:segmentation-performance} illustrates the performance of the
segmentation approach in classifying student responses. The results indicate
that a segmentation threshold of 1 yielded a surprisingly low performance across
all metrics, with 68\% agreement, a Cohen's $\kappa$ of 0.3, an F1 score of
0.61, precision of 0.76, and recall of 0.64. The model achieved its highest
performance at a threshold of 2, with 81\% agreement, a Cohen's $\kappa$ of
0.61, an F1 score of 0.80, precision of 0.83, and recall of 0.81.

These findings are somewhat counterintuitive, as we expect the presence of fewer
segments to be associated with relational responses. Upon closer examination, we
found that the model was segmenting students' descriptions of function
definitions as separate segments from the rest of the code, leading to a higher
number of segments for responses that might otherwise be considered relational.
This behavior is inconsistent with the few-shot examples provided to the model,
where the entire relational response---including the description of the function
definition---is mapped to the entire function, encompassing both its definition
and body. In contrast, for multi-structural responses, the description of the
function definition (e.g., "defines a function foo that takes...") is mapped
only to the line defining the function (e.g., \code{def foo():}). 

\paragraph{\textbf{Post-processing Improvements}}

To address this, we implemented a post-processing step that removed segments
containing \textit{only} function definitions. The results of this
post-processing step, shown in
Figure~\ref{fig:segmentation-performance-no-definitions}, align more closely
with expectations. For a threshold of 1, the percent agreement increased from
68\% to 81\%. Looking at the results for other metrics, the Cohen's $\kappa$
improved from 0.30 to 0.61, the F1 score increased from 0.61 to 0.79, precision
rose from 0.76 to 0.84, and recall improved from 0.61 to 0.80. The performance
for both a threshold of 1 and 2 were roughly equivalent.

\section{Discussion}

\begin{figure*}
  \centering
  \begin{subfigure}{0.45\textwidth}
  \begin{tikzpicture}
    \tikzstyle{promptBox} = [rectangle, draw, align=left, rounded corners, inner sep=10pt]
  
    \node at (2.5, 2) [promptBox, font=\footnotesize, text width=0.9\columnwidth,  align=left] {
        \textbf{Segmented Explanation:} [Iterates through an array] [until it finds an even number] and [adds it to a variable x]. [Return x at the end].
    };

    \foreach \x in {1,2,...,5} {
        \draw[fill=gray!50, draw=black] (\x-1,0) rectangle (\x,0.5);
    }
    
    \foreach \x in {1,2,3,4} {
        \draw[fill=blue!10!purple!50, draw=black] (\x-1,0) rectangle (\x,0.5);
    }
    
    \node at (0, -0.3) [anchor=north] {\textbf{Relational}};
    \node at (5, -0.3) [anchor=north] {\textbf{Multistructural}};
    
    \foreach \x/\label in {2/{1}, 3/{2}, 4/{3}, 5/{4}, 6/{5}} {
        \node at (\x-1,.5) [font=\footnotesize, anchor=south] {\label};
    }
\end{tikzpicture}

    \caption{An example of a segmentation feedback bar for a function that contains 5 lines and, therefore, a maximum of 5 segments. The number of purple blocks indicates how many segments were parsed and thus how close an answer is to being relational, with an ideal relational answer having a single segment (one purple block) that represents the entire function.}
    \label{fig:segment-feedback}
  \end{subfigure}
  \hfill
  \begin{subfigure}{0.45\textwidth}
    \centering
    \begin{tikzpicture}[
        node distance=.5cm and 1.5cm,
        prompt/.style={rectangle, draw=black, fill=blue!10!purple!20, rounded corners, text width=3cm, align=left, font=\footnotesize},
        code/.style={rectangle, draw=black, fill=white, rounded corners, text width=3cm, align=left, font=\footnotesize},
        arrow/.style={->, thick, >=Stealth}
    ]
    
    \tikzstyle{promptBox} = [rectangle, draw, align=left, rounded corners, inner sep=10pt]
    
    \node (explain) [promptBox, font=\footnotesize, text width=0.9\columnwidth,  align=left] {
        \textbf{Explanation:} Iterates through an array until it finds an even number and adds it to a variable x. Return x at the end.
    };
    
    \node[prompt, fill=viridis1!50, below of=explain, yshift=-.8cm, xshift=-2.3cm] (prompt1) {Iterates through an array };
    \node[prompt, below=of prompt1,fill=viridis2!50] (prompt2) {until it finds an even number};
    \node[prompt, below=of prompt2,fill=viridis3!50] (prompt3) {adds it to a variable \texttt{x}.};
    \node[prompt, below=of prompt3,fill=viridis4!50] (prompt4) {Return \texttt{x} at the end.};
    
    \node[code, right=of prompt1, fill=viridis1!50] (code1) {\texttt{for (int i = 0; i < size; i++) \{}};
    \node[code, right=of prompt2, fill=viridis2!50] (code2) {\texttt{if (arr[i] \% 2 == 0) \{}};
    \node[code, right=of prompt3, fill=viridis3!50] (code3) {\texttt{x += arr[i];}};
    \node[code, right=of prompt4, fill=viridis4!50] (code4) {\texttt{return x;}};
    
    \draw[arrow] (prompt1.east) -- (code1.west);
    \draw[arrow] (prompt2.east) -- (code2.west);
    \draw[arrow] (prompt3.east) -- (code3.west);
    \draw[arrow] (prompt4.east) -- (code4.west);
    
    \end{tikzpicture}
  \caption{Visualization mapping segments to their associated code.}
  \label{fig:prompt-segmentation-visualization}
  \end{subfigure}
  \vspace{-2mm}
  \caption{Potential student-facing feedback mechanisms for use with prompt segmentation classification}\label{fig:semechanismsgmentation-performance-comparison}
  \vspace{-2mm}
\end{figure*}

Given the reasonable performance of the segmentation approach, we are left to 
consider the application of the approach in the context of autograding and
feedback. In Section~\ref{sec:post-processing-considerations}, we reflect
on our findings regarding the improvements seen by the inclusion of the
post-processing step of removing segments that only map to the function
definitions. Following this, we discuss in Section~\ref{sec:feedback} the ways
in which this mechanism could be integrated into instruction as a feedback
mechanism for students. In both cases, we conclude each section with a summary
of the design affordances of our system that enable these features.

\subsection{Post-Processing Considerations}~\label{sec:post-processing-considerations}

Our findings with regard to the post-processing step of removing segments that
only map to the function definition highlights the importance of considering the
context in which the exercises are used when tuning the autograder. In the course
where these data were collected, students were instructed to provide instructions to
the LLM with details on the function definition to avoid ambiguity. This
highlights how post-processing steps---be they across all questions or specific
to certain questions---can be used to further refine the use of segmentation for
classifying responses as multi-structural or relational. For example, consider
the following code.

\begin{lstlisting}[
  language=C, 
  basicstyle=\ttfamily\small, 
  keywordstyle=\bfseries\color{purple}, 
  commentstyle=\itshape\color{green}
  numberstyle=\tiny, 
  stepnumber=1, 
  frame=single,
  numbersep=5pt, 
  tabsize=4
]
int returnIndexOfFirstZero(int arr[], int size) {
    for (int i = 0; i < size; i++) {
        if (arr[i] == 0) {
            return i;
        }
    }
    return -1;
}
\end{lstlisting}

A relational response might be \textit{``defines a function that takes an array
and its size, returns the index of the first zero if one exists, otherwise, it
returns -1.''} Under the segmentation approach demonstrated in this paper, this
would likely be divided into three segments: 1) the function definition, 2) the
general behavior, and 3) the default behavior. However, through post-processing,
the function definition and default behavior could be removed or accounted for
through some other means so the student is not marked down for providing a
complete description.

\vspace{2mm}
\noindent
\setlength{\fboxsep}{10pt} 
\setlength{\fboxrule}{0.5pt} 
\fcolorbox{black}{gray!10}{
\parbox{\dimexpr\linewidth-2\fboxsep-2\fboxrule\relax}{
\textbf{Design Affordances:} Our system is designed to offer flexibility through two key features: 
\setlength{\leftmargini}{15pt} 
\begin{itemize}
    \item Enabling question authors to provide few-shot examples that align their segmentation expectations.
    \item Allowing easy post-processing of the structured output to adapt it to the specific context in which the problem is being presented.
\end{itemize}
}}

\subsection{Segmentation as a Feedback Mechanism}\label{sec:feedback}

We offer two proposed use cases of how this system should be used in a student
facing fashion.

The first is simply as a mechanism for providing students with an indication of
how far they are from providing a high-level description of code
(Figure~\ref{fig:segment-feedback}). As seen from our results, particularly
after some degree of post-processing has occurred, guiding students towards
responses that contain 1-2 segments is likely to coincide with guiding them away
from multi-structural responses and towards relational ones. In addition to
aligning with the grading expectations of EiPE questions, such a feedback
mechanism could be used to give students a sense of progression rather than
simply giving them dichotomous feedback, which is often less effective than more
granular feedback~\cite{hao_towards_2022}.

The second is a mechanism for helping students bridge the gap between their
descriptions of code and how elements of those natural language descriptions
ultimately manifest as code
(Figure~\ref{fig:prompt-segmentation-visualization}). In terms of integrating
this mechanism into the ``Code Generation Based Grading'' pipeline proposed by
\citet{smith2024code}, by performing prompt segmentation \textit{in addition to}
code generation we could not only provide feedback on \textit{correctness} and
\textit{high-levelness}, but also provide visual mappings between the segments
of their prompt and their associated line(s) of code. Such feedback may be
particularly valuable for helping students debug responses that might be
categorized as \textit{multi-structural error}, where a description is
multi-structural in nature but contains incorrect descriptions of one or more
structures.

\vspace{2mm}
\noindent
\setlength{\fboxsep}{10pt} 
\setlength{\fboxrule}{0.5pt} 
\fcolorbox{black}{gray!10}{
\parbox{\dimexpr\linewidth-2\fboxsep-2\fboxrule\relax}{
\textbf{Design Affordances:} The current output of the system described in this paper can be used to provide feedback by: 
\setlength{\leftmargini}{15pt} 
\begin{itemize}
    \item Indicating to the student that they should reduce the number of highlight segments of code.
    \item Map response segments to associated code to show how portions of a description manifest as code.
\end{itemize}
}}

\section{Limitations and Future Work}

Though our results suggest the system is currently effective at performing
segmentation insofar as it can be used to differentiate between multi-structural
and relational responses, there remains the question of whether its approach to
performing segmentation aligns with human expectations. Further, formalized
analysis of the segments produced is needed before the mapping feedback shown in
Figure~\ref{fig:prompt-segmentation-visualization} can be used reliably.
Similarly, avenues for ensuring this reliability such as creating a mapping
system between natural language and code that is both more formalized and
generalizable across examples along with fine-tuning a model on such examples
may prove useful in terms of ensuring the reliability of the system. Future work
will consider such avenues towards further refining the system.

\section{Conclusion}
The segmentation-based classification method introduced in this study
demonstrates substantial promise in distinguishing between multi-structural and
relational responses to code comprehension tasks. By prompting large language
models with few-shot examples, the system achieves reasonable accuracy which was
further improved through limited post processing of the resulting segments.
Future work aims to refine the approach further through model fine-tuning and
explore the effectiveness of using the resulting segments to provide feedback to
students.

\bibliographystyle{ACM-Reference-Format}
\balance
\bibliography{sample-base}

\end{document}